\documentclass[aps,twocolumn,floatfix,superscriptaddress,showpacs,amsfonts,amssymb,amsmath]{revtex4}
\usepackage{bm}
\usepackage{epsfig}
\usepackage{color}

\newcommand{\ex}[1]{(\ref{#1})}
\newcommand{\eq}[1]{Eq.~(\ref{#1})}
\newcommand{\eqs}[2]{Eqs.~(\ref{#1}--\ref{#2})} 
\newcommand{\Eq}[1]{Eq.~(\ref{#1})}
\newcommand{\Eqs}[2]{Eqs.~(\ref{#1}--\ref{#2})} 
\newcommand{\fig}[1]{Fig.~\ref{#1}}
\newcommand{\Fig}[1]{Fig.~\ref{#1}}

\newcommand{\etal}{\textit{et al.}}

\newcommand{\vu}{\bm{u}}
\newcommand{\vB}{\bm{B}}
\newcommand{\vdel}{\bm{\nabla}}
\newcommand{\delperp}{\nabla_\perp}

\newcommand{\lt}{\left}
\newcommand{\rt}{\right}

\newcommand{\bea}{\begin{eqnarray}}
\newcommand{\eea}{\end{eqnarray}}
\newcommand\be{\begin{equation}}
\newcommand\ee{\end{equation}}

\newcommand{\gmax}{\gamma_\text{max}}
\newcommand{\kmax}{\kappa_\text{max}}

\newcommand{\Lsheet}{L_\text{CS}}
\newcommand{\deltacs}{\delta_\text{CS}}

\newcommand{\kdelta}{\kappa^2 \epsilon^2}
\newcommand{\sqkdelta}{\kappa \epsilon}

\newcommand{\DD}{\Delta'}
\newcommand{\xinf}{\xi_0}

\newcommand{\dd}{\partial}

\begin{document}


\title{Instability of current sheets and formation of plasmoid chains}
\author{N.\ F.\ Loureiro}
\affiliation{Center for Multiscale Plasma Dynamics, 
University of Maryland, College Park, Maryland 20742-3511, USA}
\affiliation{Plasma Physics Laboratory, Princeton University, Princeton, New Jersey 08543, USA}
\author{A.\ A.\ Schekochihin}
\affiliation{Blackett Laboratory, Imperial College, London~SW7~2BW, United Kingdom}
\affiliation{King's College, University of Cambridge, Cambridge CB2 1ST, United Kingdom}
\author{S.\ C.\ Cowley}
\affiliation{Department of Physics and Astronomy, UCLA, Los Angeles, California 90095-1547, USA}
\affiliation{Blackett Laboratory, Imperial College, London~SW7~2BW, United Kingdom}
\date{\today}

\begin{abstract}
Current sheets formed in magnetic reconnection events are found to 
be unstable to high-wavenumber perturbations. The instability 
is very fast: its maximum growth rate scales as 
$S^{1/4}v_A/\Lsheet$, where $\Lsheet$ is the length of the sheet, 
$v_A$ the Alfv\'en speed and $S$ the Lundquist number. As a result, 
a chain of plasmoids (secondary islands) is formed, 
whose number scales as~$S^{3/8}$. 
\end{abstract}

\pacs{52.35.Vd, 52.35.Py, 94.30.cp, 96.60.Iv}

\maketitle

\begin{figure}
\epsfig{file=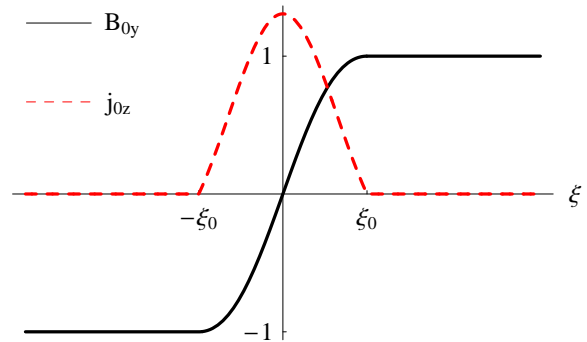,width=8cm}
\caption{\label{equilib_plot} 
(Color online) The equilibrium magnetic field $B_{0y}$ 
and the current density $j_{0z}=\dd_x B_{0y}$
(normalized to $v_A$ and $v_A/\deltacs$, respectively).}
\end{figure}

Magnetic reconnection is a plasma phenomenon 
in which oppositely directed magnetic field lines are 
driven together, break and rejoin in a topologically different configuration. 
It is an essential element in our understanding of the 
solar flares and the magnetotail,\cite{sweet_review,dungey_61,bhatta_04} 
where it is directly observed,\cite{yokoyama_01,xiao_06} 
as well as of other astrophysical plasmas. 
On Earth, it plays a crucial role in the dynamics of magnetically 
confined plasmas in fusion devices.\cite{hastie_97}\\
\indent There are two standard reconnection models: 
the Sweet-Parker (SP) model,\cite{parker,sweet} 
and the Petscheck model.\cite{pet_64} The latter 
is very appealing as it predicts fast reconnection rates 
similar to the observed ones. 
However, numerical simulations have consistently failed to 
reproduce it unless a spatially inhomogeneous 
anomalous resistivity is used\cite{malyshkin_05} 
or Hall physics is invoked.\cite{birn_01,bhatta_04} 
In contrast, the SP reconnection, characterized by long current 
sheets ($\sim$ system size) and slow reconnection rates
$\propto\eta^{1/2}$, where $\eta$ is the plasma resistivity, 
is routinely observed both in experiments\cite{yamada_00} and 
in simulations.\\
\indent The break up of current sheets and 
formation of plasmoids (secondary islands) 
appears to be a generic feature of reconnecting systems. 
Plasmoids have been observed both in solar flares\cite{lin_05} 
and in the magnetotail.\cite{zong_04} 
In numerical simulations, plasmoid formation 
has been reported in many different set ups, from fluid\cite{biskamp_86,loureiro_05} 
to fully kinetic.\cite{daughton_06,drake_plasmoids_06}
Plasmoids have been popular in theories of magnetic reconnection 
and related phenomena: e.g., 
they have been invoked as a plausible mechanism for accelerating 
reconnection, either by decreasing the effective length of the SP current sheet 
and/or by triggering anomalous-resistivity mechanisms associated with 
small-scale plasma effects;\cite{shibata_01}
a multiple-plasmoid scenario has been suggested to explain the production of energetic 
electrons during reconnection events;\cite{drake_nature_06}
it has been conjectured that periodic ejection of plasmoids in star-disk 
systems could account for the knot-like structures observed in stellar 
jets.\cite{uzdensky_04}\\
\indent A theoretical understanding of the mechanism whereby the plasmoids 
are formed has, however, been lacking. It has been believed that plasmoid 
formation is due to a standard tearing instability\cite{FKR} 
of the current sheet. Bulanov \etal\cite{bulanov_78} considered the onset 
of such an instability by 
taking into account a linear outflow along the sheet and repeating 
the standard tearing-mode calculation. While this gave a qualitative 
indication of the mildly stabilizing role of the outflow, it was necessarily 
a nonrigorous approach because the resistivity could not be neglected 
anywhere inside the current sheet and no other small parameter was available. 
In this Letter, we pursue a different strategy by  
considering a large-aspect-ratio sheet and using the inverse of its 
aspect ratio as the small parameter. We show that 
multiple plasmoids form and that, unlike in Bulanov \etal,\cite{bulanov_78}  
the instability is very fast, with a rate much larger than 
the ideal (Alfv\'enic) rate.\\
\begin{figure}
\epsfig{file=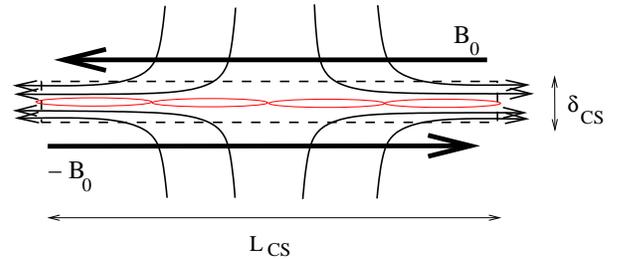,width=8cm}
\caption{\label{cartoon}(Color online) Schematic drawing of the current sheet with in- and 
outflows and a forming plasmoid chain.}
\end{figure}
%
\indent Theoretical estimates for the speed of magnetic reconnection 
in natural systems are usually based on the idea that 
a current sheet is formed (\fig{cartoon}), with the length $\Lsheet$ determined 
by the system's global properties and the width 
$\Lsheet/\sqrt{S}$, where $S=v_A\Lsheet/\eta$ 
is the Lundquist number, $v_A$ is the upstream Alfv\'en speed, 
and $\eta$ is the magnetic diffusivity. 
Let us consider such a current sheet, and ask if it can be stable
over any significant period of time. The SP reconnection 
time is $(\Lsheet/v_A)\sqrt{S}$. If we consider times 
much shorter than this time, we can determine the structure 
of the current sheet by seeking a stationary solution 
of the induction equation 
\be
\label{induction}
\dd_t\vB + \vu \cdot \vdel\vB = \vB \cdot \vdel\vu 
- \vB\vdel\cdot\vu + \eta \nabla^2\bm B,
\ee
where $\vu$ is the velocity field and $\vB$ the magnetic 
field, which we will measure in velocity units. 
While the desired resistive equilibrium is stationary, 
it is not static. 
It is a consistent feature of current sheets, 
both measured\cite{yamada_00} and simulated,\cite{uzdensky_00} 
that they support linear outflows along themselves. 
If incompressibility is assumed and 
the reconnection is considered in two dimensions 
$(x,y)$ with a current sheet along the $y$ axis, then, 
inside the current sheet, $u_x=-\Gamma_0 x$ and $u_y=\Gamma_0 y$. 
Here $\Gamma_0 = 2v_A/\Lsheet$, so the outflows are Alfv\'enic.
Obviously, this choice of $\bm u$ is a drastic simplification of the
real flow profile (ignoring, for example, the flow vorticity). It 
is only intended as a minimal model incorporating what we believe to be 
the key feature, namely the linearly increasing Alfv\'enic outflow.\\
\indent A simple equilibrium solution $\vB_0$ of \eq{induction} exists 
that accommodates this flow pattern: 
$B_{0x}=0$, and $B_{0y} = B_{0y}(x)$ satisfies 
\be
\deltacs^2\dd^2_x B_{0y}+\dd_x(x B_{0y})=0,
\ee
where $\deltacs=(\eta/\Gamma_0)^{1/2}$ is the characteristic 
width of the current sheet. 
The solution of this equation that vanishes and 
changes sign at $x=0$ (the center of the sheet) is 
$B_{0y}=v_A f(\xi)$, where $\xi=x/\deltacs$ and 
\be
\label{equilib_B}
f(\xi)= \alpha e^{-\xi^2/2}\int_0^\xi dz\,e^{z^2/2}.
\ee
The integration constant $\alpha$ is chosen by matching this solution 
with the magnetic field outside the sheet: 
$B_{0y}=\pm v_A$, which is a solution of \eq{induction} 
with constant inflows $u_x=\mp u_0$, $u_y=0$. 
In order to complete our simple model of the current sheet, 
we must choose a suitable point $x=\pm x_0$ 
at which the inside and outside solutions 
could be matched. The flow is discontinuous at 
this point: $\vu=(u_0,0)$ for $x<-x_0$,  
$\vu=(-\Gamma_0 x,\Gamma_0 y)$ for $-x_0<x<x_0$ 
and $\vu=(-u_0,0)$ for $x>x_0$, where $u_0=\Gamma_0 x_0$. 
We consider this to be an acceptable simplification 
of the real, more complicated, and continuous, flow profile. 
The natural matching point is the point where $B_{0y}(x)$ 
has its maximum (minimum), $B_{0y}(\pm x_0)=0$, and where, 
therefore, the current vanishes. 
This gives $x_0=\xinf\deltacs$, where $\xinf\approx1.31$. 
We then require $f(\pm\xinf)=\pm1$, so $\alpha=\xinf$. 
The equilibrium magnetic field and current are plotted 
in \fig{equilib_plot}.\\ 
%
\indent We shall show that this equilibrium is subject to 
a very fast linear instability. We consider the two-dimensional 
case and solve the Reduced MHD equations.\cite{strauss_76} 
Denoting $\lt\{\phi,\psi\rt\}\equiv\partial_{x}\phi\,\partial_{y}\psi-
\partial_{y}\phi\,\partial_{x}\psi$, we~have
\bea
\label{RMHD_vort}
&&\dd_t\delperp^2\phi + \lt\{\phi,\delperp^2\phi\rt\} =
\lt\{\psi,\delperp^2\psi\rt\},\\
\label{RMHD_psi}
&&\dd_t\psi + \lt\{\phi,\psi\rt\} = 
\eta\delperp^2\psi + E_0.
\eea
Here $\phi$ and $\psi$ are the stream and flux functions 
of the in-plane velocity and magnetic field, so 
$\vu=(-\dd_y\phi,\dd_x\phi)$, $\vB=(-\dd_y\psi,\dd_x\psi)$. 
\Eq{RMHD_vort} is the curl of the (inviscid) 
equation for an incompressible conducting fluid, 
\eq{RMHD_psi} is the induction equation \ex{induction} 
uncurled. $E_0$ is the equilibrium electric field, 
which must satisfy $\dd_x E_0=0$. The model of 
equilibrium flows described above corresponds 
to $\phi_0=\Gamma_0 xy$ for $|x|<x_0$ and 
$\phi_0=\pm\Gamma_0 x_0 y$ for $|x|>x_0$, which 
satisfies \eq{RMHD_vort}. The magnetic-field profile 
\ex{equilib_B} satisfies \eq{RMHD_psi}, provided we choose 
$E_0=-v_A\Gamma_0\deltacs\alpha$ for $|x|<x_0$ 
and $E_0=-v_A\Gamma_0 x_0$ for $|x|>x_0$.\\ 
\indent Let us consider small perturbations to this equilibrium, so 
$\psi=\psi_0+\delta\psi$, $\phi=\phi_0+\delta\phi$, and linearize 
\eqs{RMHD_vort}{RMHD_psi}. If we seek solutions in the form 
$\delta\phi(x,y,t)=\phi_1(x,t)\exp[ik(t)y]$ and  
$\delta\psi(x,y,t)=\psi_1(x,t)\exp[ik(t)y]$, 
where $k(t)=k_0\exp(-\Gamma_0 t)$, then 
$\phi_1$ and $\psi_1$ satisfy 
\bea
&&(\dd^2_x-k^2)\dd_t\phi_1 - \Gamma_0 x \dd_x(\dd^2_x-k^2)\phi_1 + 2\Gamma_0 k^2\phi_1
\nonumber\\
\label{phi_lin}
&&\qquad\qquad=\lt[B_{0y}(x)(\dd^2_x-k^2)-B''_{0y}(x)\rt]ik\psi_1,\\
\label{psi_lin}
&&\dd_t \psi_1 - \Gamma_0 x \dd_x\psi_1 - B_{0y}(x)ik\phi_1 = \eta(\dd^2_x-k^2)\psi_1.
\eea
We now seek exponentially growing solutions, 
$\phi_1(x,t)=-i\Phi(x)\exp(\gamma t)$ and $\psi_1(x,t)=\Psi(x)\exp(\gamma t)$. 
A solvable eigenvalue problem for $\gamma$ can be obtained 
if we assume $\gamma\gg\Gamma_0$, so the terms proportional to 
$\Gamma_0$ can be neglected and $k(t)\approx k_0$. 
Note that the presence of the 
linear in- and outflows in this approximation is only felt 
via the equilibrium profile $B_{0y}(x)$. 
Rewriting \eqs{phi_lin}{psi_lin} in terms of the dimensionless 
variable $\xi=x/\deltacs$ and denoting 
$\kappa = k_0 v_A/\Gamma_0 = k_0\Lsheet/2$, 
$\epsilon = (\eta\Gamma_0)^{1/2}/v_A = 2\deltacs/\Lsheet$, 
and $\lambda = \gamma/\Gamma_0\kappa$, we get 
\bea
\label{Phi_eq}
&&\lambda(\Phi''-\kdelta\Phi) 
= - f(\xi)(\Psi''-\kdelta\Psi) + f''(\xi)\Psi,\\
\label{Psi_eq}
&&\lambda\Psi - f(\xi)\Phi = {1\over\kappa}(\Psi'' - \kdelta\Psi),
\eea
where the derivatives are with respect to $\xi$.\\ 
\indent The eigenvalue problem given by 
\eqs{Phi_eq}{Psi_eq} is mathematically similar to the 
standard tearing mode problem,\cite{FKR} except 
the role of resistivity is played by $1/\kappa$. 
We shall assume that this parameter is small, i.e., we shall 
look for high-wavenumber perturbations of the current sheet. 
Another small parameter is $\epsilon=(2/S)^{1/2}$, 
which is the inverse aspect ratio of the sheet. 
We shall assume that $\sqkdelta\ll1$. All these assumptions 
will prove to be correct for the fastest growing modes.\\ 
\indent We now proceed as in the standard tearing mode calculation, 
considering first the outer region, $\xi\sim1$, and then 
the inner region, $\xi\ll1$.\\
\indent{\em Outer region.} The behavior here is ``ideal.'' 
Assuming that $(1/\kappa)\ll\lambda\ll1$, we get from  
\eq{Psi_eq} $\Phi=\lambda\Psi/f(\xi)$ and then from 
\eq{Phi_eq}
\be
\label{outer_Psi}
\Psi''=\left[\frac{f''(\xi)}{f(\xi)} +\kdelta\right]\Psi.
\ee
Given the functional form of the equilibrium $f(\xi)$, solving this equation 
exactly is difficult. Instead, we shall solve it perturbatively, 
using $\kdelta\ll1$. Neglecting $\kdelta$ to lowest order, 
we have an equation whose one solution is $f(\xi)$. It is then 
easy to find the second solution, so the general solution can be 
written as follows 
\be
\label{Psi_gen}
\Psi^\pm(\xi)=C_1^\pm f(\xi)+C_2^\pm f(\xi)\int_{\pm \xinf}^\xi\frac{dz}{f^2(z)},
\ee
where $C_1^\pm$ and $C_2^\pm$ are constants of integration and 
$\pm$ refers to the solution at positive and negative values of $\xi$.
We ask for the solution (but not its derivative) to be continuous 
at $\xi=0$. At small $\xi$, we have $f(\xi)\approx\alpha\xi$, so 
the integral in \eq{Psi_gen} is dominated by its upper limit 
and we find that $C_2^+=C_2^-=-\alpha\Psi(0)$.\\  
\indent The constants $C_1^\pm$ are found by matching the solution \ex{Psi_gen} 
to the outer solution at $|\xi|>\xinf$ (outside the current sheet). 
There we have, instead of \eq{outer_Psi},
\be
\label{cowley}
\Psi''=\kdelta\Psi, 
\ee	
so $\kdelta$ can no longer be neglected. 
The solution that decays at $\xi\to\pm\infty$ is 
$\Psi^\pm= C_3^\pm\exp(\mp\sqkdelta\xi)$. Matching 
this solution and its derivative to the solution \ex{Psi_gen}
and using $f(\pm\xinf)=\pm1$ and $f'(\pm\xinf)=0$, 
we get $C_2^\pm = \pm\alpha\Psi(0)/\sqkdelta$ 
and $C_3^\pm = \alpha\Psi(0)\exp(\sqkdelta\xinf)/\sqkdelta$. 
Thus, for $|\xi|<\xinf$, 
\be
\label{Psi_sol}
\Psi^\pm(\xi) = \pm{\alpha\Psi(0)\over\sqkdelta}\,f(\xi)
- \alpha\Psi(0)f(\xi)\int_{\pm\xinf}^\xi\frac{dz}{f^2(\xi)}, 
\ee
and for $|\xi|>\xinf$, 
\be
\Psi^\pm(\xi) = {\alpha\Psi(0)\over\sqkdelta}\,\exp\lt[\sqkdelta(\xinf\mp\xi)\rt].
\ee
\begin{figure}
\epsfig{file=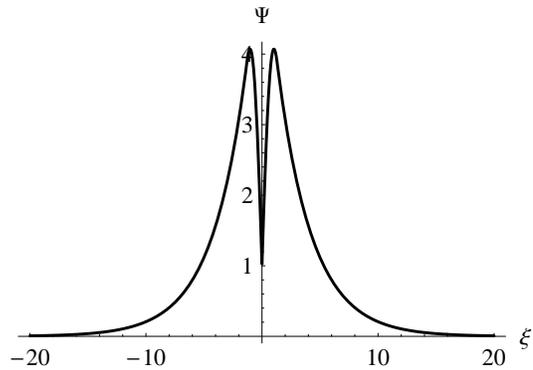,width=8cm}
\caption{\label{Psi_plot} The outer solution evaluated for 
$\epsilon^{-1}=10^3$ at $\kappa\approx340$, 
corresponding to the maximum growth rate.}
\end{figure}  
%
This outer solution is plotted in \fig{Psi_plot}. It has 
a discontinuous derivative at $\xi=0$ and is a classic 
unstable eigenfunction known from the tearing-mode problem. 
The instability parameter $\DD=[\Psi'(+0)-\Psi'(-0)]/\Psi(0)$~is
\be
\label{dprime}
\DD=\frac{2\alpha^2}{\sqkdelta} 
+ \alpha^2\left[\int_{-\xinf}^{\xinf}\frac{dz}{f^2(z)}\right],
\ee
where $[\dots]$ denotes the nonsingular part of the integral. 
The second term in \eq{dprime}, which is equal to $0.57$, 
can be dropped compared to the first, 
so to lowest order we have simply $\DD=2\alpha^2/\sqkdelta$.\footnote{Going 
to the next order in our outer solution 
results in an additional subdominant contribution to $\DD$ comparable 
to the second term in \eq{dprime}. Note that 
while the contribution of the second term in \eq{Psi_sol} 
to $\DD$ is negligible, the term itself cannot be dropped 
because it ensures that the solution has a finite value at $\xi=0$. 
Note also that there exists an unstable solution that is entirely confined 
inside the current sheet, so that $\Psi(\pm\xinf)=0$. For this solution, 
$C_1^\pm=0$ in \eq{Psi_gen} and to the lowest order in $\kdelta$, 
$\DD$ is given just by the second term in \eq{dprime}. Solving 
\eq{outer_Psi} to the next order in $\kdelta$, we find 
$\DD(\kappa)\approx 0.57 - 1.47 \kdelta$.
The growth rate then is obtained from \eq{disp_rel} in the limit $\Lambda\ll1$. 
The result is $\gamma/\Gamma_0\approx 0.61\,\kappa^{2/5}\DD(\kappa)^{4/5}$,
whence $\kmax\approx0.28\epsilon^{-1}$ and $\gmax\sim\epsilon^{-2/5}$. 
We will not consider this solution because it grows 
slower than the unconfined mode and its validity 
hinges on the numerical smallness of $\kmax^2\epsilon^2\approx 0.08$.}\\ 
\indent{\em Inner region.} Here $\xi\ll1$ and we can assume $f(\xi)\approx\alpha\xi$.
\Eqs{Phi_eq}{Psi_eq}, assuming $\dd_\xi\gg1$, become 
\bea
&&\lambda\Phi'' = - \alpha\xi\Psi'',\\
&&\lambda\Psi - \alpha\xi\Phi = {1\over\kappa}\Psi''.
\eea
Since $\DD=2\alpha^2/\sqkdelta\gg1$, this eigenvalue problem is 
mathematically similar to the one solved by Coppi~\etal\cite{coppi_76} 
for large-$\DD$ tearing modes. It reduces to solving the 
following transcendental equation for the growth rate 
\be
\label{disp_rel}
-\frac{\pi}{8}\,(\kappa\alpha )^{1/3}\Lambda^{5/4}\frac
{\Gamma\lt((\Lambda^{3/2}-1)/4\rt)}
{\Gamma\lt((\Lambda^{3/2}+5)/4\rt)} 
= \DD = \frac{2\alpha^2}{\sqkdelta},
\ee
where $\Gamma$ is the gamma function
and $\Lambda=\lambda\alpha^{-2/3}\kappa^{1/3}$. 
The width of the inner region is given by 
$\delta=(\lambda/\kappa)^{1/4}\alpha^{-1/2}$.
Recall that $\lambda=\gamma/\Gamma_0\kappa$
and $\kappa=k_0 v_A/\Gamma_0$.\\ 
\indent \Eq{disp_rel} has two interesting limits: 
assuming~$\Lambda\ll1$,
\be
\label{FKR_scaling}
\gamma/\Gamma_0\approx1.63\,\kappa^{-2/5}\epsilon^{-4/5};
\ee
assuming $\Lambda \to 1-$ (from below),
\be
\label{coppi_scaling}
\gamma/\Gamma_0\approx(\alpha\kappa)^{2/3}-\frac{\sqrt{\pi}}{3\alpha}\kappa^2\epsilon.
\ee
The maximum growth rate $\gmax$ lies between these two asymptotics. 
Comparing the two terms in \eq{coppi_scaling} shows 
that it is attained for $\kmax\sim\epsilon^{-3/4}\gg1$ and, therefore,  
$\gmax/\Gamma_0\sim\epsilon^{-1/2}\gg1$. 
We note also that $\kmax\epsilon\sim\epsilon^{1/4}$, 
$\lambda\sim\epsilon^{1/4}$ and the inner layer width $\delta\sim\epsilon^{1/4}$
for the fastest growing mode. 
This confirms all of the ordering assumptions we made in our calculation.\\ 
\begin{figure}
\epsfig{file=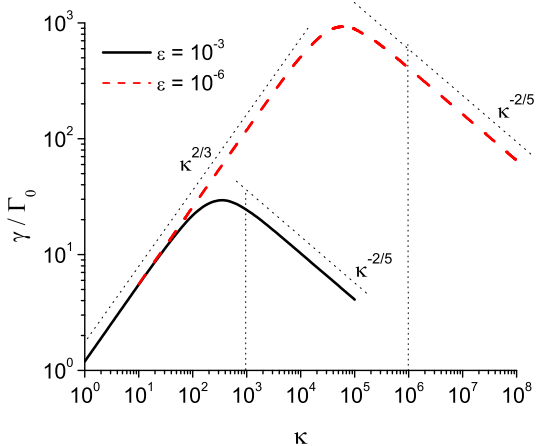,width=8cm} 
\caption{\label{gamma_plots} 
(Color online) The growth rate calculated from \eq{disp_rel}  
for two values of the aspect ratio: $\epsilon^{-1}=10^3$ and $10^{6}$.} 
\end{figure}
%
\indent \Fig{gamma_plots} shows the dependence $\gamma(\kappa)$ resulting from the numerical 
solution of \eq{disp_rel} for two different values of the current sheet aspect 
ratio $\epsilon^{-1}$. Both scalings \ex{FKR_scaling} and \ex{coppi_scaling} 
are manifest. 
The vertical lines identify the values of $\kappa$ for which
$\kappa \epsilon=1$. Strictly speaking, our calculation is
only valid for values of $\kappa$ significantly to the
left of this line.
This is, however, of secondary importance, as the maximum of the 
growth rate lies well within the region of validity of our asymptotic ordering 
for $\epsilon^{-1}\gtrsim10^3$.\\
\indent We have shown analytically that SP current sheets with 
large aspect ratios are intrinsically unstable to high-wave-number 
perturbations. Since $\epsilon^{-1}=(S/2)^{1/2}$, the maximum growth rate scales 
with the Lundquist number as $\gmax\sim S^{1/4}\Gamma_0$. 
Typical Lundquist numbers in plasmas of interest are extremely large 
(e.g., $S\approx10^{12}$ in the solar corona), so the 
instability of the current sheet is extremely fast compared to the 
Alfv\'en time $\sim\Gamma_0^{-1}$. One immediate consequence of this 
is that stable current sheets with aspect ratios above some 
critical value cannot exist. Numerical 
simulations\cite{biskamp_86,loureiro_05} suggest that this value is $\sim10^2$,
corresponding to $S_c\sim10^4$. Above this value, 
the sheet breaks up and a chain of plasmoids is formed. 
Their number scales as $\kmax\sim S^{3/8}$ and their length is 
$\sim S^{1/8}$ larger than the current-sheet width $\deltacs$.\\
\indent We note that, as far as we know, no numerical simulations of current sheets 
with aspect ratios significantly exceeding $\sim10^2$ have been reported. 
At these values, our asymptotic theory is not yet rigorously applicable, so 
we cannot test it against the existing simulations of SP reconnection.
A nonasymptotic extrapolation of our results suggests that only a very small
number of plasmoids is to be expected --- these are, indeed, 
seen.\cite{biskamp_86,loureiro_05,samtaney_05,knoll_chacon_06,daughton_06,drake_plasmoids_06}\\ 
\indent Since the growth rate we have found 
is much larger than the inverse Alfv\'en time $v_A/\Lsheet$, the plasmoid
width should become comparable to the inner-layer width 
$\delta\sim S^{-1/8}\deltacs$ before the plasmoids can be expelled 
from the current sheet by the Alfv\'enic outflows. 
The plasmoid evolution then becomes nonlinear. 
In order to have a quantitative theory of how the plasmoids 
affect the reconnection, one needs to understand what happens in the 
nonlinear regime. The dynamics in this regime 
will be dictated by a competition between three processes: 
the nonlinear growth and saturation of the plasmoids due to reconnection, 
the plasmoid coalescence,\cite{finn_and_kaw,biskamp_82} 
and the expulsion of the plasmoids along the current sheet by the Alfv\'enic 
outflows. Numerical results on the stalling of the coalescence instability at large 
$S$ suggest that multiple plasmoids can survive in the nonlinear 
regime rather than coalescing into a single plasmoid.\cite{knoll_chacon_06} The 
formation of multiple large plasmoids has also been reported.\cite{samtaney_05,drake_plasmoids_06} 
If the width of the plasmoids can indeed become bigger than the width of 
the current sheet before they coalesce or are expelled, estimates of the 
SP reconnection time must no longer be based on the parameters of the original 
current sheet but rather on some effective reconnection region whose width is 
determined by the saturated plasmoid chain. 
A nonlinear study of the current sheet instability and plasmoid
dynamics, as well as comparing the theory proposed in this
Letter against more sophisticated and richer physics models, can only
be performed by means of numerical simulations.
We are presently developing a computer code tailored to the study of this
problem. Results will be presented in a separate publication.

\begin{acknowledgments}
Discussions with D.~A.~Uzdensky are gratefully acknowledged. 
N.F.L.\ was supported by the Center for Multiscale Plasma Dynamics, 
DOE Fusion Science Center Cooperative Agreement ER54785.
He thanks the Leverhulme International Network for 
Magnetised Plasma Turbulence for travel support. 
A.A.S.\ was supported by an STFC Advanced Fellowship. 
\end{acknowledgments}


%
%
%
%

\end{document}